\documentclass[prl,twocolumn,showpacs,preprintnumbers,amsmath,amssymb,superscriptaddress]{revtex4-1}

\usepackage{graphicx}
\usepackage{dcolumn}
\usepackage{bm}
\usepackage{units}

\begin{document}

\title{Critical behavior of nanocrystalline gadolinium: Evidence for a new universality class}

\author{A.~Ferdinand}
\affiliation{Experimentalphysik, Universit\"at des Saarlandes, Postfach 151150, D-66041 Saarbr\"ucken, Germany}
\author{A.-C.~Probst}
\affiliation{Experimentalphysik, Universit\"at des Saarlandes, Postfach 151150, D-66041 Saarbr\"ucken, Germany}
\author{A.~Michels}
\email[]{andreas.michels@uni.lu}
\affiliation{Physics and Materials Science Research Unit, University of Luxembourg, 162A~Avenue de la Fa\"iencerie, L-1511 Luxembourg, Grand Duchy of Luxembourg}
\author{R.~Birringer}
\email[]{r.birringer@nano.uni-saarland.de}
\affiliation{Experimentalphysik, Universit\"at des Saarlandes, Postfach 151150, D-66041 Saarbr\"ucken, Germany}
\author{S.~N.~Kaul}
\email[]{kaul.sn@gmail.com}
\affiliation{School of Physics, University of Hyderabad, Central University P.O., Hyderabad-500 046, India}

\date{\today}

\begin{abstract}
We report on how nanocrystal size affects the critical behavior of the rare-earth metal Gd near the ferromagnetic-to-paramagnetic phase transition. The asymptotic critical behavior of the coarse-grained polycrystalline sample (with an average crystallite size of $L \cong \unit[100]{\mu m}$) is that of a (pure) \textsl{uniaxial dipolar} ferromagnet, as is the case with single-crystal Gd, albeit the width of the asymptotic critical region (ACR) is reduced. As the grain size approaches $\sim \unit[30]{nm}$, the ACR is so narrow that it could not be accessed in the present experiments. Inaccessibly narrow ACR for $L \sim \unit[30]{nm}$ and the continuous increase in the width of ACR as $L$ decreases from $\unit[16]{nm}$ to $\unit[9.5]{nm}$ basically reflects a crossover to the \textsl{random uniaxial dipolar} fixed point caused by the quenched random-exchange disorder prevalent at the internal interfaces (grain boundaries).
\end{abstract}


\maketitle

For a long time, the rare-earth metal Gd was considered to be an archetypal isotropic three-dimensional ($d = 3$) Heisenberg ferromagnet for twofold reasons. First, magnetocrystalline anisotropy (MCA) is expected to be extremely weak \cite{legvold80} since Gd is made up of pure $S$-state Gd$^{3+}$ ions with $L = 0$. Second, isotropic Rudermann-Kittel-Kasuya-Yosida (RKKY) interactions, which mimic the Heisenberg form, couple the magnetic moments localized at the sites of the hexagonal closed-packed (hcp) lattice of Gd. During the past decade, theoretical efforts to understand the physical origin of \textsl{observed} MCA \cite{geldart1987a, tosti03prl}, ferromagnetic ground state \cite{blugel2003, nolting2004} and unusually large volume magnetostriction near the Curie temperature, $T_C$, \cite{mohn2004} have resulted in substantial modifications to the long-held simplistic picture ($d = 3$ Heisenberg ferromagnet) of Gd. Consequently, the significant contribution of long-range dipole-dipole interactions to MCA as well as to polarizing the localized $4f$ spins, and the polarization of the $5spd$ and $6s$ conduction-band electron spins due to RKKY coupling to the localized $4f$ spins is now well established. Another important recent development is the resolution \cite{kaul2000} of the basic issue of whether Gd is a ferromagnet with a collinear spin structure or an antiferromagnet with a helical spin arrangement, akin to other heavy rare-earth metals at temperatures ranging between the spin-reorientation temperature $T_{\mathrm{SR}} = 230 \, \mathrm{K}$ and $T_C = 292.77 \, \mathrm{K}$ \cite{coey1999,kaul2000}.

The long-standing (spanning nearly four decades) controversy \cite{kaulspringer} about the asymptotic critical behavior of Gd near the ferromagnetic (FM) to paramagnetic (PM) phase transition has finally been put to rest by demonstrating \cite{kaul1999a,kaul1999b} that single power laws alone cannot adequately describe the observed temperature variations of spontaneous magnetization, $M(T, 0)$, and intrinsic susceptibility, $\chi(T)$, in the asymptotic critical region (ACR), but do so only when the multiplicative logarithmic corrections to these power laws, predicted by RG calculations for a $d = 3$ uniaxial dipolar ferromagnet \cite{frey90,RiedMillevUDFM}, are taken into account. To be more specific, zero-field electrical resistivity/specific heat, $C_{H = 0}$, \cite{geldart1987b,bednarz1993}, $M(T, 0)$ and $\chi(T)$, taken along the $c$-axis (easy direction of magnetization) of a high-purity Gd single crystal, respectively follow the RG-predicted temperature variations, $C_{H = 0} \sim \left| \varepsilon \right|^{-\alpha} \left| \ln \varepsilon \right|^{1/3}$, with $\varepsilon = (T - T_C) / T_C$ and $\alpha = 0$, $M(T, 0) \sim (-\varepsilon)^{\beta} \left| \ln |\varepsilon| \right|^{1/3}$, with $\beta = 0.5$, and
\begin{equation}
\label{PUD}
\chi^{-1}(T) = \Gamma_p^{-1} \, \varepsilon^{\gamma} \, \left| \ln\varepsilon \right|^{-x} \,\,,
\end{equation}
over two decades in reduced temperature. For instance, Eq.~(\ref{PUD}) is obeyed in the ACR $5.1 \times 10^{-5} \leq \varepsilon \leq 2.05 \times 10^{-3}$ for $\varepsilon > 0$, with $T_C = \unit[292.77(1)]{K}$, $\gamma = 1.0008(5)$, and $x = 0.329(1)$ \cite{kaul1999a,kaul1999b}. Thus, Gd (single crystal) belongs to the $d = 3$ uniaxial dipolar universality class and will henceforth be referred to as the pure uniaxial dipolar (PUD) ferromagnet. In essence, experiment (theory) has so far employed single crystal hcp Gd metal (idealized imperfection-free model systems).

A field of growing research activity relates to the influence of quenched randomness, i.e., compositional, topological, bond- and/or site-disorder, as well as confinement, proximity, and symmetry breaking on magnetic properties \cite{farle1993,gajdzik1998,dami02,had2003,jackson05,kruk2006,michels08epl,dobrichprb2012,Andrianov2013}. Remarkable effects are routinely observed in real magnetic materials where free surfaces, internal interfaces, line and point defects, local composition fluctuations, and preparation history form integral parts of the material's microstructure. The effects of randomness on second order phase transitions have been a subject of sustained interest \cite{fisher1968,lubensky1975,Korzhenevskii1,Korzhenevskii2,DudkaFolk,RAMNewLTPhase,UnusualCritBehav,sandvik2010}. In this context, the celebrated ``Harris criterion'' \cite{HarrisKriterium} provides general guidelines, i.e., (i) the addition of short-range disorder to a non-random (pure) system, which undergoes a second-order phase transition, should not affect the sharpness of the transition (hence leave the critical exponents unaltered), if the specific-heat critical exponent of the pure system is  $\alpha_p < 0 $, and (ii) a crossover to a new type of (random) critical behavior could occur, if $\alpha_p > 0$. Since the specific heat of pure $d = 3$ uniaxial dipolar ferromagnet (e.g., Gd) \textsl{diverges asymptotically} as $\left| \ln \varepsilon \right|^{1/3}$, according to the Harris criterion (ii) \cite{HarrisKriterium}, adding quenched randomness/disorder to PUD ferromagnet should result in a new type of asymptotic critical behavior. Indeed, RG calculations  \cite{DilutedUDFM,DilutedUDFM5}, based on the quenched-random exchange Ising model (which includes both quenched site- and bond-diluted models) with dipolar interactions, yield the multiplicative corrections to the leading singular behavior in susceptibility and specific heat at $T_C$ that are drastically different from their PUD counterparts. For instance, they predict the intrinsic susceptibility for $\varepsilon > 0$ as
\begin{equation}
\label{RUD}
\chi^{-1}(T) = \Gamma_r^{-1} \, \varepsilon^{\gamma} \, \exp\left(-\sqrt{D \, \left| \ln\varepsilon \right|} \right) \,\,,
\end{equation}
with the mean-field value of $\gamma = 1$ and a universal constant $D \cong 0.113$ \cite{DilutedUDFM5}. The asymptotic critical behavior, now characterized by the \textsl{new} correction term $\exp(- \sqrt{D \, |\ln\varepsilon|} )$, has been assigned a new universality class, namely, the random uniaxial dipolar (RUD) universality class.

The main objective of the present study is to ascertain whether or not the quenched randomness/disorder present in nanocrystalline (NC) Gd metal gives rise to deviations from the PUD behavior and if so, can the asymptotic leading singularity be identified and a universality class assigned to a possible new fixed point?

Before presenting the experimental details, we shortly digress into specifying the type of randomness/disorder in NC Gd. It is a polycrystalline aggregate made up of randomly oriented nanometer-sized grains embedded in a manifold of grain boundaries (GBs). The core region of such GBs accommodates the atomic mismatch between adjacent but differently oriented nanocrystallites. The atomic site mismatch in the GBs gives rise to random site disorder that translates into random exchange interaction, whereas the uniaxial anisotropy axis, oriented along the $c$-axis of hcp crystal structure in each individual nanocrystal, varies randomly from grain to grain leading to a random distribution of easy axes in the grain ensemble. The structural correlation length of such nanocrystallites, i.e., the grain size, $L$, can be varied depending on preparation and subsequent annealing conditions. NC Gd may, thus, be considered as a model system to study the influence of random site disorder and randomness in uniaxial anisotropy on the FM-PM phase transition. The control parameter $L$ permits manipulation of the strength of quenched random site disorder as well as the degree of randomness of anisotropy. Actually, these quantities are coupled, since both scale as $L^{-1}$ \cite{lcomment}.

Several plausible scenarios for phase transition in NC Gd that can be invoked are: (A) The specific type of randomness and disorder present in NC Gd has no effect on the PUD universality class, but only the non-universal quantities ($T_C$ and critical amplitudes) get altered. (B) The transition is smeared \cite{Korzhenevskii1,Korzhenevskii2,DudkaFolk,RAMNewLTPhase,UnusualCritBehav}. (C) If the random anisotropy dominates the critical behavior, a crossover from the (pure) uniaxial dipolar fixed point to an isotropic dipolar fixed point may occur. (D) If, on the other hand, the quenched random site disorder, prevalent at grain surfaces/interfaces and in the core regions of GBs, controls the asymptotic critical behavior, NC Gd should behave as a RUD ferromagnet in the ACR. Furthermore, this scenario becomes more probable as $L$ reduces.

\begin{figure}
\includegraphics[width=0.80\columnwidth]{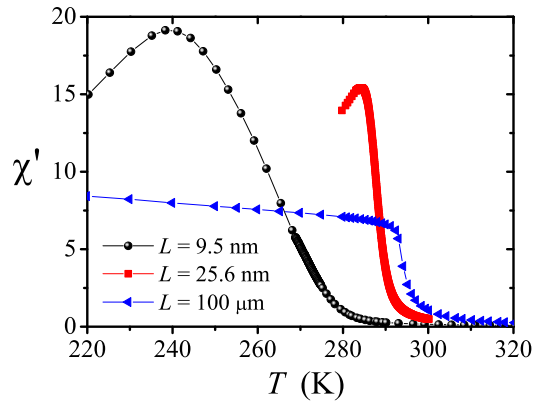}
\caption{\label{fig1} (Color online) Real part of the ac-susceptibility of different nanocrystalline Gd samples as a function of temperature.}\vspace{-0.25cm}
\end{figure}

In order to unravel the asymptotic critical behavior of NC Gd, ac-susceptibility, $\chi_{\mathrm{ac}}$, was measured as a function of temperature in steps of $\unit[20 - 40]{mK}$, particularly in the critical region, at the ac driving field of typical amplitude $\unit[1]{Oe}$ and frequency $\unit[100]{Hz}$ on several NC Gd samples of different grain sizes, using Quantum Design PPMS magnetometer. Details of sample preparation and characterization are furnished in \cite{dami02,dobrichprb2012}. Figure~\ref{fig1} depicts the real part of $\chi_{\mathrm{ac}}(T)$, $\chi^{\prime}(T)$, for three representative samples with a view to highlight the relevance of grain size to the magnetic behavior. After correcting $\chi^{\prime}(T)$ for demagnetization to arrive at the intrinsic susceptibility, $\chi(T)$, the so-called ``range-of-fit'' (ROF) analysis, detailed in \cite{kaul1999a,kaul1999b}, is used to determine the effective (denoted by the subscript ``$\mathrm{eff}$'') and asymptotic amplitudes and critical exponents, appearing respectively in the single power law (SPL), $\chi^{-1}(T) = \Gamma_{\mathrm{eff}}^{-1} \, \varepsilon^{\gamma_{\mathrm{eff}}}$, and Eqs.~(\ref{PUD}) and (\ref{RUD}).

\begin{figure}
\includegraphics[width=0.55\columnwidth]{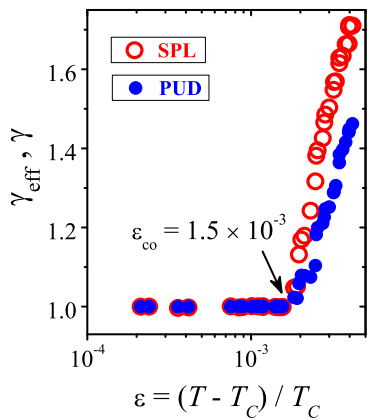}
\caption{\label{fig2} (Color online) Temperature variations of the effective, $\gamma_{\mathrm{eff}}$, and asymptotic, $\gamma$, susceptibility critical exponents for coarse-grained Gd with $L = \unit[100]{\mu m}$, deduced from the ROF analysis based on the SPL and PUD [Eq.~(\ref{PUD})] expressions, respectively, with $T_C$ fixed at $\unit[291.917]{K}$.}\vspace{-0.25cm}
\end{figure}

\begin{figure}
\includegraphics[width=0.78\columnwidth]{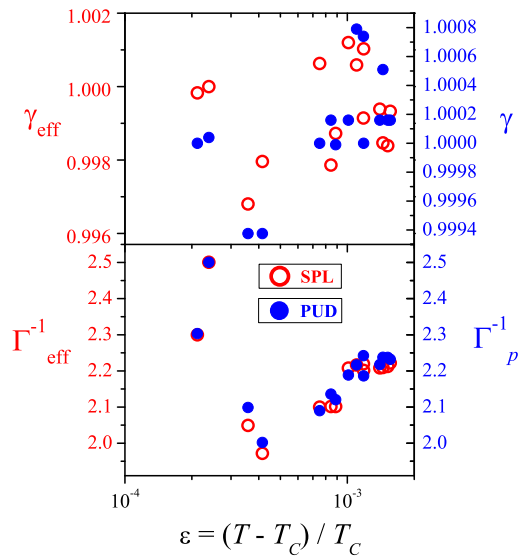}
\caption{\label{fig3} (Color online) Temperature variations of the effective, $\gamma_{\mathrm{eff}}(\varepsilon)$, and asymptotic, $\gamma(\varepsilon)$, critical exponents (top panel) and of the corresponding inverse critical amplitudes $\Gamma^{-1}_{\mathrm{eff}}(\varepsilon)$ and $\Gamma^{-1}_p(\varepsilon)$ (bottom panel) for coarse-grained Gd ($L = \unit[100]{\mu m}$) in the asymptotic critical regime ($\varepsilon < \varepsilon_{\mathrm{co}}$).}\vspace{-0.25cm}
\end{figure}

Figure~\ref{fig2} displays the temperature variations of the effective, $\gamma_{\mathrm{eff}}$, and asymptotic, $\gamma $, critical exponents for susceptibility, obtained from the ROF analysis \cite{kaul1999a,kaul1999b}, based on SPL and PUD expressions, for the coarse-grained ($L = \unit[100]{\mu m}$) Gd sample. Note that the logarithmic correction exponent $x$ in Eq.~(\ref{PUD}) is kept constant at the RG value $x = 1/3$ in the ROF analysis, which yields the same value $T_C = \unit [291.917(3)]{K}$ for $T_C$, within the uncertainty limits, in both SPL and PUD cases. From the data presented in Fig.~\ref{fig2} it is evident that the exponent $\gamma_{\mathrm{eff}}$ (SPL) as well as $\gamma$ (PUD) are very close to the mean-field value of $1$ for temperatures up to a well-defined crossover (``$\mathrm{co}$'') temperature, $\varepsilon_{\mathrm{co}} = 1.5 \times 10^{-3}$, beyond which they increase steeply. The temperature range $0 < \varepsilon < \varepsilon_{\mathrm{co}}$ equals the width of the ACR. The observation that $\gamma_{\mathrm{eff}} \rightarrow 1$ as $\varepsilon \rightarrow 0$ is a strong indication of the (pure) uniaxial dipolar behavior in the ACR. Further support for this inference comes from the following result. Consistent with the RG prediction that, in the asymptotic limit $\varepsilon \rightarrow 0$, critical amplitudes and exponents should attain constant values, e.g., $\gamma = 1.0$ in the PUD case, $\Gamma^{-1}_p$ and $\gamma = 1.0001(7)$ have \textsl{less scatter} (Fig.~\ref{fig3}) than $\Gamma^{-1}_{\mathrm{eff}}$ and $\gamma_{\mathrm{eff}} = 0.999(3)$ within the ACR.

\begin{figure}
\includegraphics[width=0.55\columnwidth]{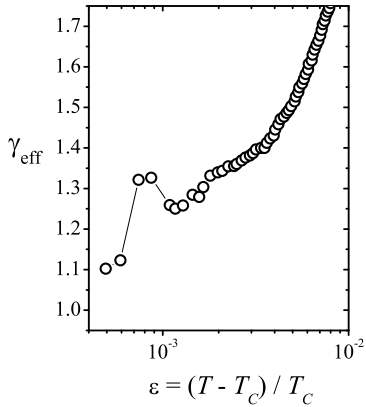}
\caption{\label{fig4} Temperature variation of the effective susceptibility critical exponent $\gamma_{\mathrm{eff}}(\varepsilon)$ for NC Gd ($L = \unit[25.6]{nm}$), obtained from the ROF analysis based on the single power law.}\vspace{-0.25cm}
\end{figure}

As evidenced from the results of the SPL-ROF analysis displayed in Fig.~\ref{fig4} (cf.~Fig.~\ref{fig2}), $\gamma_{\mathrm{eff}}(\varepsilon)$ exhibits a completely different behavior in the NC Gd sample with $L = \unit[25.6]{nm}$ as compared to coarse-grained Gd. For this sample, there is no clear indication of an asymptotic critical behavior. Instead, $\gamma_{\mathrm{eff}}$ tends to approach $1$ when $\varepsilon < 4.9 \times 10^{-4}$ (the temperature closest to $T_{C} = \unit[285.63]{K}$ in the experiment) and attains the value $\gamma_{\mathrm{eff}} = 1.35(5)$ on either side (i.e., in the temperature ranges $7.4 \times 10^{-4} \leq \varepsilon \leq 8.7 \times 10^{-4}$ and $1.8 \times 10^{-3} \leq \varepsilon \leq 3.5 \times 10^{-3}$) of the minimum [$\gamma_{\mathrm{eff}}(\varepsilon_{\mathrm{min}}) = 1.25(5)$] occurring at  $\varepsilon_{\mathrm{min}} = 1.17 \times 10^{-3}$. Such non-monotonous temperature variation of $\gamma_{\mathrm{eff}}$ is indicative of a series of crossovers in the critical region, which result from an interplay between the different types of interactions.

For a PUD ferromagnet, the RG calculations \cite{frey90,RiedMillevUDFM} predict the sequence of crossovers uniaxial dipolar (UD) $\longrightarrow$ isotropic dipolar (ID) $\longrightarrow$ isotropic short-range Heisenberg (IH) $\longrightarrow$ Gaussian regime, as the temperature increases from $T_{C}$. According to the RG treatment \cite{bruce77,frey91} of ferromagnets with isotropic short-range Heisenberg and long-range dipolar interactions, the characteristic experimental signature \cite{srinath2000} for the ID-IH crossover is a well-defined minimum in $\gamma_{\mathrm{eff}}(\varepsilon)$ (at $\varepsilon_{\mathrm{dip}}$ with $\gamma_{\mathrm{eff}}(\varepsilon_{\mathrm{dip}}) \simeq 1.28$) that separates the asymptotic ID regime (characterized by the critical exponent $\gamma^{\mathrm{ID}} = 1.372$) from the IH regime (with $\gamma^{\mathrm{IH}} = 1.365$). A direct comparison between theory and experiment, thus, reveals that in the present experiments on the sample with $L = \unit[25.6]{nm}$, the asymptotic PUD regime could not be accessed as it is extremely narrow and lies well below $\varepsilon = 4.9 \times 10^{-4}$; otherwise, the observed temperature variation of $\gamma_{\mathrm{eff}}$ conforms well with the RG predictions. The behavior of $\gamma_{\mathrm{eff}}(\varepsilon)$ similar to that found in the sample $L = \unit[25.6]{nm}$ is also observed in $L = \unit[34]{nm}$ (data not shown) and even in the latter case, PUD ACR remained inaccessible although the $T_{C} = \unit[287.22]{K}$ was approached as closely as $\varepsilon = 4.7 \times 10^{-4}$.

With reference to the coarse-grained specimen, the behavior of $L = \unit[25.6]{nm}$ seems to suggest that the reduction in the grain size by $4$ orders of magnitude weakens the effective uniaxial anisotropy to some extent, but promotes the quenched random-exchange disorder (QRD) at grain surfaces/interfaces and in GBs, such that, at such grain sizes, $L \cong \unit[25 - 34]{nm}$, QRD has a strength just sufficient to qualify as a relevant perturbation (or a relevant scaling field in the RG sense) which, in turn, renders the PUD fixed point unstable and causes a crossover to the RUD fixed point. ACR is thus so narrow as to remain inaccessible to experiments. If this line of argument is pursued, the RUD ACR is expected to progressively increase in width as a result of increasing QRD with decreasing $L$. Consistent with this expectation, we observe that the width of the RUD ACR increases continuously as the grain size reduces from $L = \unit[16]{nm}$ (data not shown) to $L = \unit[9.5]{nm}$; the upper bound of the RUD ACR increases from $ \varepsilon_{\mathrm{co}} = 2.5 \times 10^{-3}$ with $T_{C} = \unit[285.481]{K}$ for $L = \unit[16]{nm}$ to $ \varepsilon_{\mathrm{co}} = 8.6 \times 10^{-3}$ with $T_{C} = \unit[252.335]{K}$ for $L = \unit[9.5]{nm}$.

\begin{figure}
\includegraphics[width=0.55\columnwidth]{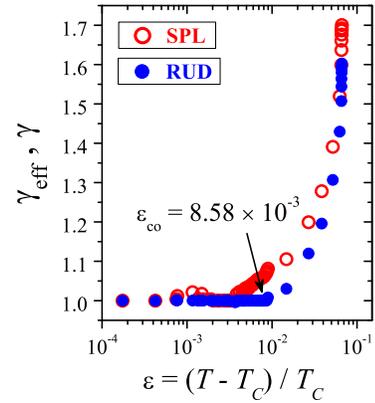}
\caption{\label{fig5} (Color online) Temperature variations of the effective, $\gamma_{\mathrm{eff}}$, and asymptotic, $\gamma$, susceptibility critical exponents for the NC Gd sample with $L = \unit[9.5]{nm}$, deduced from the ROF analysis based on the SPL and RUD [Eq.~(\ref{RUD})] expressions, respectively, with $T_C$ fixed at $\unit[252.335]{K}$.}
\end{figure}

\begin{figure}
\includegraphics[width=0.82\columnwidth]{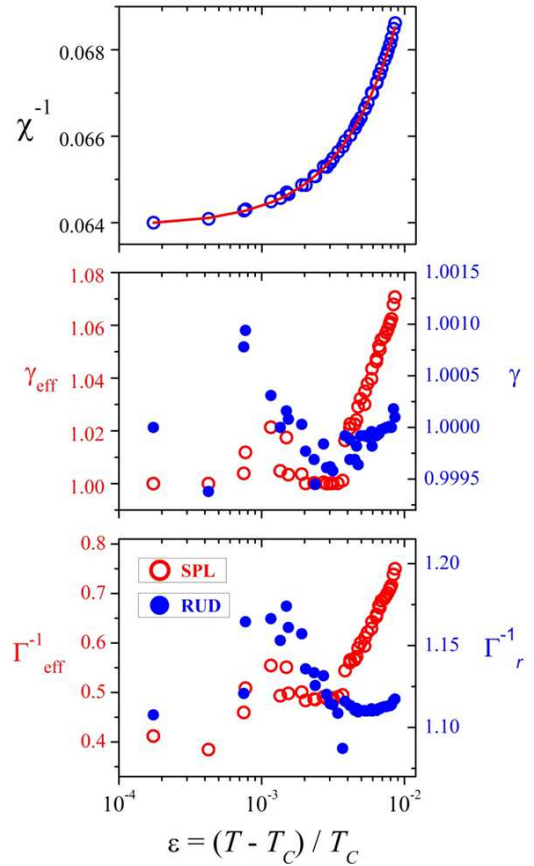}
\caption{\label{fig6} (Color online) Top panel: The best RUD fit (continuous curve) to the inverse intrinsic susceptibility, $\chi^{-1}(T)$, of NC Gd with $L = \unit[9.5]{nm}$ in the ACR ($\varepsilon < \varepsilon_{\mathrm{co}}$). Temperature variations of the effective and asymptotic critical exponents $\gamma_{\mathrm{eff}}(\varepsilon)$ and $\gamma(\varepsilon)$ (middle panel), and of the corresponding inverse critical amplitudes $\Gamma^{-1}_{\mathrm{eff}}(\varepsilon)$ and $\Gamma^{-1}_r(\varepsilon)$ (bottom panel) in the ACR.}
\end{figure}

Figure~\ref{fig5} displays the temperature variations of $\gamma_{\mathrm{eff}}$ and $\gamma$ that the ROF analysis, based on the SPL and the RUD [Eq.~(\ref{RUD})] expressions, yields for the NC Gd sample with $L = \unit[9.5]{nm}$, when $T_C$ is fixed at $\unit[252.335]{K}$. Judging by the reduced sum of deviation squares, we find that the SPL does not describe $\chi^{-1}(T)$ as accurately as the RUD expression [Eq.~(\ref{RUD})] in the ACR (Fig.~\ref{fig6}, top panel). The middle and bottom panels of Fig.~\ref{fig6} show the results of the ROF analysis of the $\chi^{-1}(T)$ data (the top panel), based on the SPL (the \textsl{effective} critical amplitude, $\Gamma_{\mathrm{eff}}^{-1}$, and critical exponent, $\gamma_{\mathrm{eff}}$) and the RG RUD (the \textsl{asymptotic} critical amplitude, $\Gamma^{-1}_r$, and critical exponent, $\gamma$) expressions. Inclusion of the multiplicative logarithmic correction, besides the leading single power law, i.e., Eq.~(\ref{RUD}), vastly improves the robustness of the fitting parameters against the variation in the temperature range of the fit (ACR being the largest fit range); e.g., compare $\Gamma_{\mathrm{eff}}^{-1} = 0.55(20)$ with $\Gamma^{-1}_r = 1.13(4)$ and $\gamma_{\mathrm{eff}} = 1.035(35)$ with $\gamma = 1.0002(8)$. But for the change in the ACR width with $L$, these results are representative of the samples with $L = \unit[12]{nm}$ and $L = \unit[16]{nm}$ as well (data not shown). In stark contrast to the extremely narrow ACR in the $L = \unit[34]{nm}$ and $L = \unit[25.6]{nm}$ samples, the ACR widens at smaller grain sizes. A plausible explanation for this observation has already been provided.

For the discussion of our results, we treat the single-crystalline (SC) Gd metal as the reference state with regard to structure and FM-PM phase transition. The change of microstructure from single-crystalline to coarse-grained polycrystalline Gd with a grain size of $\unit[100]{\mu m}$ leads to a slight narrowing of the ACR from $\varepsilon_{\mathrm{co}} = 2.05 \times 10^{-3}$ to $\varepsilon_{\mathrm{co}} = 1.5 \times 10^{-3}$, but leaves the PUD asymptotic critical behavior unaltered. Assuming that the spin-spin correlation length $\xi$ (the distance over which the order-parameter fluctuations are correlated) grows well beyond $L$ due to sufficiently strong spin coupling across the GBs but does not reach the system size at $T = T_{C}$ (as contrasted with SC Gd, wherein $\xi$ diverges at $T_{C}$), the effective uniaxial anisotropy weakens due to the averaging over the random crystal orientations within the spin-correlated volume ($\sim \xi^3$) with the result that the ACR of the UD ferromagnet shrinks. However, as $L$ approaches the nanometer range, the number of atoms (and hence spins) at the grain surface increases rapidly at the cost of the atoms within the core. Consequently, QRD picks up in strength and, beyond a threshold, causes a crossover from the PUD to RUD asymptotic critical behavior. In NC Gd, this threshold is reached at $L \cong \unit[34]{nm}$ and the RUD behavior in the ACR is found in the samples with  $L \cong \unit[9.5 - 16]{nm}$ (Figs.~\ref{fig5} and \ref{fig6}).

In summary, an elaborate analysis of the intrinsic magnetic susceptibility reveals that, as is the case with single-crystalline Gd, the asymptotic critical behavior of polycrystalline coarse-grained Gd (grain size: $\sim \unit[100]{\mu m}$) in the critical region near the paramagnetic-to-ferromagnetic phase transition is that of a (pure) \textsl{uniaxial dipolar} ferromagnet. For nanocrystalline Gd with a grain size of $\sim \unit[30]{nm}$, asymptotic critical behavior could not be discerned, which we believe is due to a crossover to the random uniaxial dipolar fixed point with an extremely narrow critical region. At grain sizes $\lesssim \unit[16]{nm}$, nanocrystalline Gd behaves as a \textsl{random uniaxial dipolar} ferromagnet in the asymptotic critical region and it belongs to the random uniaxial dipolar universality class.

This work was supported under the DFG Project No.~MI 738/3-2 and by the National Research Fund of Luxembourg (ATTRACT Project No.~FNR/A09/01). We thank Dr.~Shinto~P.~Mathew for assisting in the data analysis.

\bibliographystyle{apsrev}

\begin{thebibliography}{40}
\expandafter\ifx\csname natexlab\endcsname\relax\def\natexlab#1{#1}\fi
\expandafter\ifx\csname bibnamefont\endcsname\relax
  \def\bibnamefont#1{#1}\fi
\expandafter\ifx\csname bibfnamefont\endcsname\relax
  \def\bibfnamefont#1{#1}\fi
\expandafter\ifx\csname citenamefont\endcsname\relax
  \def\citenamefont#1{#1}\fi
\expandafter\ifx\csname url\endcsname\relax
  \def\url#1{\texttt{#1}}\fi
\expandafter\ifx\csname urlprefix\endcsname\relax\def\urlprefix{URL }\fi
\providecommand{\bibinfo}[2]{#2}
\providecommand{\eprint}[2][]{\url{#2}}

\bibitem[{\citenamefont{Legvold}(1980)}]{legvold80}
\bibinfo{author}{\bibfnamefont{S.}~\bibnamefont{Legvold}}, in
  \emph{\bibinfo{booktitle}{Ferromagnetic Materials}}, edited by
  \bibinfo{editor}{\bibfnamefont{E.~P.} \bibnamefont{Wohlfarth}}
  (\bibinfo{publisher}{North-Holland Publishing Company},
  \bibinfo{address}{Amsterdam}, \bibinfo{year}{1980}),
  vol.~\bibinfo{volume}{1}, pp. \bibinfo{pages}{183--295}.

\bibitem[{\citenamefont{Fujiki et~al.}(1987)\citenamefont{Fujiki, De'Bell, and
  Geldart}}]{geldart1987a}
\bibinfo{author}{\bibfnamefont{N.~M.} \bibnamefont{Fujiki}},
  \bibinfo{author}{\bibfnamefont{K.}~\bibnamefont{De'Bell}}, \bibnamefont{and}
  \bibinfo{author}{\bibfnamefont{D.~J.~W.} \bibnamefont{Geldart}},
  \bibinfo{journal}{Phys. Rev. B} \textbf{\bibinfo{volume}{36}},
  \bibinfo{pages}{8512} (\bibinfo{year}{1987}).

\bibitem[{\citenamefont{Colarieti-Tosti
  et~al.}(2003)\citenamefont{Colarieti-Tosti, Simak, Ahuja, Nordstr\"{o}m,
  Eriksson, $\mathrm{{\AA}}$berg, Edvardsson, and Brooks}}]{tosti03prl}
\bibinfo{author}{\bibfnamefont{M.}~\bibnamefont{Colarieti-Tosti}},
  \bibinfo{author}{\bibfnamefont{S.~I.} \bibnamefont{Simak}},
  \bibinfo{author}{\bibfnamefont{R.}~\bibnamefont{Ahuja}},
  \bibinfo{author}{\bibfnamefont{L.}~\bibnamefont{Nordstr\"{o}m}},
  \bibinfo{author}{\bibfnamefont{O.}~\bibnamefont{Eriksson}},
  \bibinfo{author}{\bibfnamefont{D.}~\bibnamefont{$\mathrm{{\AA}}$berg}},
  \bibinfo{author}{\bibfnamefont{S.}~\bibnamefont{Edvardsson}},
  \bibnamefont{and} \bibinfo{author}{\bibfnamefont{M.~S.~S.}
  \bibnamefont{Brooks}}, \bibinfo{journal}{Phys. Rev. Lett.}
  \textbf{\bibinfo{volume}{91}}, \bibinfo{pages}{157201}
  (\bibinfo{year}{2003}).

\bibitem[{\citenamefont{Turek et~al.}(2003)\citenamefont{Turek,
  Kudrnovsk$\mathrm{\acute{y}}$, Bihlmayer, and Bl\"ugel}}]{blugel2003}
\bibinfo{author}{\bibfnamefont{I.}~\bibnamefont{Turek}},
  \bibinfo{author}{\bibfnamefont{J.}~\bibnamefont{Kudrnovsk$\mathrm{\acute{y}}%
$}}, \bibinfo{author}{\bibfnamefont{G.}~\bibnamefont{Bihlmayer}},
  \bibnamefont{and} \bibinfo{author}{\bibfnamefont{S.}~\bibnamefont{Bl\"ugel}},
  \bibinfo{journal}{J. Phys.: Condens. Matter} \textbf{\bibinfo{volume}{15}},
  \bibinfo{pages}{2771} (\bibinfo{year}{2003}).

\bibitem[{\citenamefont{Santos et~al.}(2004)\citenamefont{Santos, Nolting, and
  Eyert}}]{nolting2004}
\bibinfo{author}{\bibfnamefont{C.}~\bibnamefont{Santos}},
  \bibinfo{author}{\bibfnamefont{W.}~\bibnamefont{Nolting}}, \bibnamefont{and}
  \bibinfo{author}{\bibfnamefont{V.}~\bibnamefont{Eyert}},
  \bibinfo{journal}{Phys. Rev. B} \textbf{\bibinfo{volume}{69}},
  \bibinfo{pages}{214412} (\bibinfo{year}{2004}).

\bibitem[{\citenamefont{Khmelevskyi et~al.}(2004)\citenamefont{Khmelevskyi,
  Turek, and Mohn}}]{mohn2004}
\bibinfo{author}{\bibfnamefont{S.}~\bibnamefont{Khmelevskyi}},
  \bibinfo{author}{\bibfnamefont{I.}~\bibnamefont{Turek}}, \bibnamefont{and}
  \bibinfo{author}{\bibfnamefont{P.}~\bibnamefont{Mohn}},
  \bibinfo{journal}{Phys. Rev. B} \textbf{\bibinfo{volume}{70}},
  \bibinfo{pages}{132401} (\bibinfo{year}{2004}).

\bibitem[{\citenamefont{Kaul and Srinath}(2000)}]{kaul2000}
\bibinfo{author}{\bibfnamefont{S.~N.} \bibnamefont{Kaul}} \bibnamefont{and}
  \bibinfo{author}{\bibfnamefont{S.}~\bibnamefont{Srinath}},
  \bibinfo{journal}{Phys. Rev. B} \textbf{\bibinfo{volume}{62}},
  \bibinfo{pages}{1114} (\bibinfo{year}{2000}).

\bibitem[{\citenamefont{Coey et~al.}(1999)\citenamefont{Coey, Skumryev, and
  Gallagher}}]{coey1999}
\bibinfo{author}{\bibfnamefont{J.~M.~D.} \bibnamefont{Coey}},
  \bibinfo{author}{\bibfnamefont{V.}~\bibnamefont{Skumryev}}, \bibnamefont{and}
  \bibinfo{author}{\bibfnamefont{K.}~\bibnamefont{Gallagher}},
  \bibinfo{journal}{Nature} \textbf{\bibinfo{volume}{401}}, \bibinfo{pages}{35}
  (\bibinfo{year}{1999}).

\bibitem[{\citenamefont{Kaul}(2005)}]{kaulspringer}
\bibinfo{author}{\bibfnamefont{S.~N.} \bibnamefont{Kaul}}, in
  \emph{\bibinfo{booktitle}{Local-moment ferromagnets: unique properties for
  modern applications (Lecture Notes in Physics)}}, edited by
  \bibinfo{editor}{\bibfnamefont{M.}~\bibnamefont{Donath}} \bibnamefont{and}
  \bibinfo{editor}{\bibfnamefont{W.}~\bibnamefont{Nolting}}
  (\bibinfo{publisher}{Springer}, \bibinfo{address}{Berlin},
  \bibinfo{year}{2005}), pp. \bibinfo{pages}{11--30}.

\bibitem[{\citenamefont{Srinath et~al.}(1999)\citenamefont{Srinath, Kaul, and
  Kronm\"uller}}]{kaul1999a}
\bibinfo{author}{\bibfnamefont{S.}~\bibnamefont{Srinath}},
  \bibinfo{author}{\bibfnamefont{S.~N.} \bibnamefont{Kaul}}, \bibnamefont{and}
  \bibinfo{author}{\bibfnamefont{H.}~\bibnamefont{Kronm\"uller}},
  \bibinfo{journal}{Phys. Rev. B} \textbf{\bibinfo{volume}{59}},
  \bibinfo{pages}{1145} (\bibinfo{year}{1999}).

\bibitem[{\citenamefont{Srinath and Kaul}(1999)}]{kaul1999b}
\bibinfo{author}{\bibfnamefont{S.}~\bibnamefont{Srinath}} \bibnamefont{and}
  \bibinfo{author}{\bibfnamefont{S.~N.} \bibnamefont{Kaul}},
  \bibinfo{journal}{Phys. Rev. B} \textbf{\bibinfo{volume}{60}},
  \bibinfo{pages}{12166} (\bibinfo{year}{1999}).

\bibitem[{\citenamefont{Frey and Schwabl}(1990)}]{frey90}
\bibinfo{author}{\bibfnamefont{E.}~\bibnamefont{Frey}} \bibnamefont{and}
  \bibinfo{author}{\bibfnamefont{F.}~\bibnamefont{Schwabl}},
  \bibinfo{journal}{Phys. Rev. B} \textbf{\bibinfo{volume}{42}},
  \bibinfo{pages}{8261} (\bibinfo{year}{1990}).

\bibitem[{\citenamefont{Ried et~al.}(1995)\citenamefont{Ried, Millev, F\"ahnle,
  and Kronm\"uller}}]{RiedMillevUDFM}
\bibinfo{author}{\bibfnamefont{K.}~\bibnamefont{Ried}},
  \bibinfo{author}{\bibfnamefont{Y.}~\bibnamefont{Millev}},
  \bibinfo{author}{\bibfnamefont{M.}~\bibnamefont{F\"ahnle}}, \bibnamefont{and}
  \bibinfo{author}{\bibfnamefont{H.}~\bibnamefont{Kronm\"uller}},
  \bibinfo{journal}{Phys. Rev. B} \textbf{\bibinfo{volume}{51}},
  \bibinfo{pages}{15229} (\bibinfo{year}{1995}).

\bibitem[{\citenamefont{Geldart et~al.}(1987)\citenamefont{Geldart, De'Bell,
  Cook, and Laubitz}}]{geldart1987b}
\bibinfo{author}{\bibfnamefont{D.~J.~W.} \bibnamefont{Geldart}},
  \bibinfo{author}{\bibfnamefont{K.}~\bibnamefont{De'Bell}},
  \bibinfo{author}{\bibfnamefont{J.}~\bibnamefont{Cook}}, \bibnamefont{and}
  \bibinfo{author}{\bibfnamefont{M.~J.} \bibnamefont{Laubitz}},
  \bibinfo{journal}{Phys. Rev. B} \textbf{\bibinfo{volume}{35}},
  \bibinfo{pages}{8876} (\bibinfo{year}{1987}).

\bibitem[{\citenamefont{Bednarz et~al.}(1993)\citenamefont{Bednarz, Geldart,
  and White}}]{bednarz1993}
\bibinfo{author}{\bibfnamefont{G.}~\bibnamefont{Bednarz}},
  \bibinfo{author}{\bibfnamefont{D.~J.~W.} \bibnamefont{Geldart}},
  \bibnamefont{and} \bibinfo{author}{\bibfnamefont{M.~A.} \bibnamefont{White}},
  \bibinfo{journal}{Phys. Rev. B} \textbf{\bibinfo{volume}{47}},
  \bibinfo{pages}{14247} (\bibinfo{year}{1993}).

\bibitem[{\citenamefont{Farle et~al.}(1993)\citenamefont{Farle, Baberschke,
  Stetter, Aspelmeier, and Gerhardter}}]{farle1993}
\bibinfo{author}{\bibfnamefont{M.}~\bibnamefont{Farle}},
  \bibinfo{author}{\bibfnamefont{K.}~\bibnamefont{Baberschke}},
  \bibinfo{author}{\bibfnamefont{U.}~\bibnamefont{Stetter}},
  \bibinfo{author}{\bibfnamefont{A.}~\bibnamefont{Aspelmeier}},
  \bibnamefont{and}
  \bibinfo{author}{\bibfnamefont{F.}~\bibnamefont{Gerhardter}},
  \bibinfo{journal}{Phys. Rev. B} \textbf{\bibinfo{volume}{47}},
  \bibinfo{pages}{11571} (\bibinfo{year}{1993}).

\bibitem[{\citenamefont{Gajdzik et~al.}(1998)\citenamefont{Gajdzik, Trappmann,
  S\"urgers, and v.~L\"ohneysen}}]{gajdzik1998}
\bibinfo{author}{\bibfnamefont{M.}~\bibnamefont{Gajdzik}},
  \bibinfo{author}{\bibfnamefont{T.}~\bibnamefont{Trappmann}},
  \bibinfo{author}{\bibfnamefont{C.}~\bibnamefont{S\"urgers}},
  \bibnamefont{and}
  \bibinfo{author}{\bibfnamefont{H.}~\bibnamefont{v.~L\"ohneysen}},
  \bibinfo{journal}{Phys. Rev. B} \textbf{\bibinfo{volume}{57}},
  \bibinfo{pages}{3525} (\bibinfo{year}{1998}).

\bibitem[{\citenamefont{Michels et~al.}(2002)\citenamefont{Michels, Krill~III,
  and Birringer}}]{dami02}
\bibinfo{author}{\bibfnamefont{D.}~\bibnamefont{Michels}},
  \bibinfo{author}{\bibfnamefont{C.~E.} \bibnamefont{Krill~III}},
  \bibnamefont{and}
  \bibinfo{author}{\bibfnamefont{R.}~\bibnamefont{Birringer}},
  \bibinfo{journal}{J. Magn. Magn. Mater.} \textbf{\bibinfo{volume}{250}},
  \bibinfo{pages}{203} (\bibinfo{year}{2002}).

\bibitem[{\citenamefont{Yan et~al.}(2003)\citenamefont{Yan, Huang, Zhang,
  Okumura, Xiao, Stoyanov, Skumryev, Hadjipanayis, and Nelson}}]{had2003}
\bibinfo{author}{\bibfnamefont{Z.~C.} \bibnamefont{Yan}},
  \bibinfo{author}{\bibfnamefont{Y.~H.} \bibnamefont{Huang}},
  \bibinfo{author}{\bibfnamefont{Y.}~\bibnamefont{Zhang}},
  \bibinfo{author}{\bibfnamefont{H.}~\bibnamefont{Okumura}},
  \bibinfo{author}{\bibfnamefont{J.~Q.} \bibnamefont{Xiao}},
  \bibinfo{author}{\bibfnamefont{S.}~\bibnamefont{Stoyanov}},
  \bibinfo{author}{\bibfnamefont{V.}~\bibnamefont{Skumryev}},
  \bibinfo{author}{\bibfnamefont{G.~C.} \bibnamefont{Hadjipanayis}},
  \bibnamefont{and} \bibinfo{author}{\bibfnamefont{C.}~\bibnamefont{Nelson}},
  \bibinfo{journal}{Phys. Rev. B} \textbf{\bibinfo{volume}{67}},
  \bibinfo{pages}{054403} (\bibinfo{year}{2003}).

\bibitem[{\citenamefont{Jackson et~al.}(2005)\citenamefont{Jackson, Malba,
  Weir, Baker, and Vohra}}]{jackson05}
\bibinfo{author}{\bibfnamefont{D.~D.} \bibnamefont{Jackson}},
  \bibinfo{author}{\bibfnamefont{V.}~\bibnamefont{Malba}},
  \bibinfo{author}{\bibfnamefont{S.~T.} \bibnamefont{Weir}},
  \bibinfo{author}{\bibfnamefont{P.~A.} \bibnamefont{Baker}}, \bibnamefont{and}
  \bibinfo{author}{\bibfnamefont{Y.~K.} \bibnamefont{Vohra}},
  \bibinfo{journal}{Phys. Rev. B} \textbf{\bibinfo{volume}{71}},
  \bibinfo{pages}{184416} (\bibinfo{year}{2005}).

\bibitem[{\citenamefont{Kruk et~al.}(2006)\citenamefont{Kruk, Ghafari, Hahn,
  Michels, Birringer, Krill~III, Kmiec, and Marszalek}}]{kruk2006}
\bibinfo{author}{\bibfnamefont{R.}~\bibnamefont{Kruk}},
  \bibinfo{author}{\bibfnamefont{M.}~\bibnamefont{Ghafari}},
  \bibinfo{author}{\bibfnamefont{H.}~\bibnamefont{Hahn}},
  \bibinfo{author}{\bibfnamefont{D.}~\bibnamefont{Michels}},
  \bibinfo{author}{\bibfnamefont{R.}~\bibnamefont{Birringer}},
  \bibinfo{author}{\bibfnamefont{C.~E.} \bibnamefont{Krill~III}},
  \bibinfo{author}{\bibfnamefont{R.}~\bibnamefont{Kmiec}}, \bibnamefont{and}
  \bibinfo{author}{\bibfnamefont{M.}~\bibnamefont{Marszalek}},
  \bibinfo{journal}{Phys. Rev. B} \textbf{\bibinfo{volume}{73}},
  \bibinfo{pages}{054420} (\bibinfo{year}{2006}).

\bibitem[{\citenamefont{Michels et~al.}(2008)\citenamefont{Michels, D\"obrich,
  Elmas, Ferdinand, Markmann, Sharp, Eckerlebe, Kohlbrecher, and
  Birringer}}]{michels08epl}
\bibinfo{author}{\bibfnamefont{A.}~\bibnamefont{Michels}},
  \bibinfo{author}{\bibfnamefont{F.}~\bibnamefont{D\"obrich}},
  \bibinfo{author}{\bibfnamefont{M.}~\bibnamefont{Elmas}},
  \bibinfo{author}{\bibfnamefont{A.}~\bibnamefont{Ferdinand}},
  \bibinfo{author}{\bibfnamefont{J.}~\bibnamefont{Markmann}},
  \bibinfo{author}{\bibfnamefont{M.}~\bibnamefont{Sharp}},
  \bibinfo{author}{\bibfnamefont{H.}~\bibnamefont{Eckerlebe}},
  \bibinfo{author}{\bibfnamefont{J.}~\bibnamefont{Kohlbrecher}},
  \bibnamefont{and}
  \bibinfo{author}{\bibfnamefont{R.}~\bibnamefont{Birringer}},
  \bibinfo{journal}{EPL} \textbf{\bibinfo{volume}{81}}, \bibinfo{pages}{66003}
  (\bibinfo{year}{2008}).

\bibitem[{\citenamefont{D\"obrich et~al.}(2012)\citenamefont{D\"obrich,
  Kohlbrecher, Sharp, Eckerlebe, Birringer, and Michels}}]{dobrichprb2012}
\bibinfo{author}{\bibfnamefont{F.}~\bibnamefont{D\"obrich}},
  \bibinfo{author}{\bibfnamefont{J.}~\bibnamefont{Kohlbrecher}},
  \bibinfo{author}{\bibfnamefont{M.}~\bibnamefont{Sharp}},
  \bibinfo{author}{\bibfnamefont{H.}~\bibnamefont{Eckerlebe}},
  \bibinfo{author}{\bibfnamefont{R.}~\bibnamefont{Birringer}},
  \bibnamefont{and} \bibinfo{author}{\bibfnamefont{A.}~\bibnamefont{Michels}},
  \bibinfo{journal}{Phys. Rev. B} \textbf{\bibinfo{volume}{85}},
  \bibinfo{pages}{094411} (\bibinfo{year}{2012}).

\bibitem[{\citenamefont{Andrianov and Bauer}(2013)}]{Andrianov2013}
\bibinfo{author}{\bibfnamefont{A.~V.} \bibnamefont{Andrianov}}
  \bibnamefont{and} \bibinfo{author}{\bibfnamefont{E.}~\bibnamefont{Bauer}},
  \bibinfo{journal}{EPL} \textbf{\bibinfo{volume}{102}}, \bibinfo{pages}{17011}
  (\bibinfo{year}{2013}).

\bibitem[{\citenamefont{Fisher}(1968)}]{fisher1968}
\bibinfo{author}{\bibfnamefont{M.~E.} \bibnamefont{Fisher}},
  \bibinfo{journal}{Phys. Rev.} \textbf{\bibinfo{volume}{176}},
  \bibinfo{pages}{257} (\bibinfo{year}{1968}).

\bibitem[{\citenamefont{Lubensky}(1975)}]{lubensky1975}
\bibinfo{author}{\bibfnamefont{T.~C.} \bibnamefont{Lubensky}},
  \bibinfo{journal}{Phys. Rev. B} \textbf{\bibinfo{volume}{11}},
  \bibinfo{pages}{3573} (\bibinfo{year}{1975}).

\bibitem[{\citenamefont{Korzhenevskii et~al.}(1996)\citenamefont{Korzhenevskii,
  Herrmanns, and Schirmacher}}]{Korzhenevskii1}
\bibinfo{author}{\bibfnamefont{A.~L.} \bibnamefont{Korzhenevskii}},
  \bibinfo{author}{\bibfnamefont{K.}~\bibnamefont{Herrmanns}},
  \bibnamefont{and}
  \bibinfo{author}{\bibfnamefont{W.}~\bibnamefont{Schirmacher}},
  \bibinfo{journal}{Phys. Rev. B} \textbf{\bibinfo{volume}{53}},
  \bibinfo{pages}{14834} (\bibinfo{year}{1996}).

\bibitem[{\citenamefont{Korzhenevskii et~al.}(1998)\citenamefont{Korzhenevskii,
  Heuer, and Herrmanns}}]{Korzhenevskii2}
\bibinfo{author}{\bibfnamefont{A.~L.} \bibnamefont{Korzhenevskii}},
  \bibinfo{author}{\bibfnamefont{H.-O.} \bibnamefont{Heuer}}, \bibnamefont{and}
  \bibinfo{author}{\bibfnamefont{K.}~\bibnamefont{Herrmanns}},
  \bibinfo{journal}{J. Phys. A: Math. Gen.} \textbf{\bibinfo{volume}{31}},
  \bibinfo{pages}{927} (\bibinfo{year}{1998}).

\bibitem[{\citenamefont{Dudka et~al.}(2005)\citenamefont{Dudka, Folk, and
  Holovatch}}]{DudkaFolk}
\bibinfo{author}{\bibfnamefont{M.}~\bibnamefont{Dudka}},
  \bibinfo{author}{\bibfnamefont{R.}~\bibnamefont{Folk}}, \bibnamefont{and}
  \bibinfo{author}{\bibfnamefont{Y.}~\bibnamefont{Holovatch}},
  \bibinfo{journal}{J. Magn. Magn. Mater.} \textbf{\bibinfo{volume}{294}},
  \bibinfo{pages}{305} (\bibinfo{year}{2005}).

\bibitem[{\citenamefont{Aharony}(1982)}]{RAMNewLTPhase}
\bibinfo{author}{\bibfnamefont{A.}~\bibnamefont{Aharony}}, \bibinfo{journal}{J.
  Magn. Magn. Mater.} \textbf{\bibinfo{volume}{31}}, \bibinfo{pages}{1432}
  (\bibinfo{year}{1982}).

\bibitem[{\citenamefont{Igloi et~al.}(1993)\citenamefont{Igloi, Peschel, and
  Turban}}]{UnusualCritBehav}
\bibinfo{author}{\bibfnamefont{F.}~\bibnamefont{Igloi}},
  \bibinfo{author}{\bibfnamefont{I.}~\bibnamefont{Peschel}}, \bibnamefont{and}
  \bibinfo{author}{\bibfnamefont{L.}~\bibnamefont{Turban}},
  \bibinfo{journal}{Adv. Phys.} \textbf{\bibinfo{volume}{42}},
  \bibinfo{pages}{683} (\bibinfo{year}{1993}).

\bibitem[{\citenamefont{Yao et~al.}(2010)\citenamefont{Yao, Gustafsson,
  Carlson, and Sandvik}}]{sandvik2010}
\bibinfo{author}{\bibfnamefont{D.-X.} \bibnamefont{Yao}},
  \bibinfo{author}{\bibfnamefont{J.}~\bibnamefont{Gustafsson}},
  \bibinfo{author}{\bibfnamefont{E.~W.} \bibnamefont{Carlson}},
  \bibnamefont{and} \bibinfo{author}{\bibfnamefont{A.~W.}
  \bibnamefont{Sandvik}}, \bibinfo{journal}{Phys. Rev. B}
  \textbf{\bibinfo{volume}{82}}, \bibinfo{pages}{172409}
  (\bibinfo{year}{2010}).

\bibitem[{\citenamefont{Harris}(1974)}]{HarrisKriterium}
\bibinfo{author}{\bibfnamefont{A.~B.} \bibnamefont{Harris}},
  \bibinfo{journal}{J. Phys. C: Solid State Phys.}
  \textbf{\bibinfo{volume}{7}}, \bibinfo{pages}{1671} (\bibinfo{year}{1974}).

\bibitem[{\citenamefont{Aharony}(1976)}]{DilutedUDFM}
\bibinfo{author}{\bibfnamefont{A.}~\bibnamefont{Aharony}},
  \bibinfo{journal}{Phys. Rev. B} \textbf{\bibinfo{volume}{13}},
  \bibinfo{pages}{2092} (\bibinfo{year}{1976}).

\bibitem[{\citenamefont{Schuster}(1977)}]{DilutedUDFM5}
\bibinfo{author}{\bibfnamefont{H.~G.} \bibnamefont{Schuster}},
  \bibinfo{journal}{Z. Phys. B} \textbf{\bibinfo{volume}{27}},
  \bibinfo{pages}{251} (\bibinfo{year}{1977}).

\bibitem[{lco()}]{lcomment}
\bibinfo{note}{We note that the here quoted values for $L$ represent, in fact,
  volume-weighted average grain diameters based on the assumption of a
  spherical crystallite shape \cite{krill98}.}

\bibitem[{\citenamefont{Bruce}(1977)}]{bruce77}
\bibinfo{author}{\bibfnamefont{A.~D.} \bibnamefont{Bruce}},
  \bibinfo{journal}{J. Phys. C: Solid State Phys.}
  \textbf{\bibinfo{volume}{10}}, \bibinfo{pages}{419} (\bibinfo{year}{1977}).

\bibitem[{\citenamefont{Frey and Schwabl}(1991)}]{frey91}
\bibinfo{author}{\bibfnamefont{E.}~\bibnamefont{Frey}} \bibnamefont{and}
  \bibinfo{author}{\bibfnamefont{F.}~\bibnamefont{Schwabl}},
  \bibinfo{journal}{Phys. Rev. B} \textbf{\bibinfo{volume}{43}},
  \bibinfo{pages}{833} (\bibinfo{year}{1991}).

\bibitem[{\citenamefont{Srinath et~al.}(2000)\citenamefont{Srinath, Kaul, and
  Sostarich}}]{srinath2000}
\bibinfo{author}{\bibfnamefont{S.}~\bibnamefont{Srinath}},
  \bibinfo{author}{\bibfnamefont{S.~N.} \bibnamefont{Kaul}}, \bibnamefont{and}
  \bibinfo{author}{\bibfnamefont{M.-K.} \bibnamefont{Sostarich}},
  \bibinfo{journal}{Phys. Rev. B} \textbf{\bibinfo{volume}{62}},
  \bibinfo{pages}{11649} (\bibinfo{year}{2000}).

\bibitem[{\citenamefont{Krill and Birringer}(1998)}]{krill98}
\bibinfo{author}{\bibfnamefont{C.~E.} \bibnamefont{Krill}} \bibnamefont{and}
  \bibinfo{author}{\bibfnamefont{R.}~\bibnamefont{Birringer}},
  \bibinfo{journal}{Philos. Mag. A} \textbf{\bibinfo{volume}{77}},
  \bibinfo{pages}{621} (\bibinfo{year}{1998}).

\end{thebibliography}

\end{document}